\documentclass[aps,nofootinbib,twocolumn]{revtex4}
\usepackage{float}
\usepackage{graphicx}  
\usepackage{subfig}
\usepackage{amsmath}   
\usepackage{amssymb}   
\usepackage{bm} 
\usepackage{dcolumn}
\usepackage{color}
\usepackage{mathrsfs}
\usepackage{amsfonts}
\usepackage{varioref}
\usepackage{textcomp}
\usepackage{ragged2e}   

\RequirePackage[colorlinks,citecolor=blue,urlcolor=magenta,linkcolor=blue]{hyperref}
\allowdisplaybreaks

\usepackage[nameinlink,capitalise,noabbrev]{cleveref}
\crefname{section}{Sec.}{Secs.}
\Crefname{section}{Section}{Sections}
\crefname{appendix}{Appendix}{Appendices}
\Crefname{appendix}{Appendix}{Appendices}

\crefname{section}{Sec.}{Secs.}
\Crefname{section}{Section}{Sections}
\crefname{appendix}{Appendix}{Appendices}
\Crefname{appendix}{Appendix}{Appendices}

\crefformat{equation}{Eq.~(#2#1#3)}
\Crefformat{equation}{Eq.~(#2#1#3)}
\crefformat{figure}{Fig.~#2#1#3}
\Crefformat{figure}{Fig.~#2#1#3}
\crefformat{table}{Tab.~#2#1#3}
\Crefformat{table}{Tab.~#2#1#3}

\begin{document}

\title{Plunge spectra as discriminators of black hole mimickers}

\author{Sreejith Nair}
\email{sreejithnair@iitgn.ac.in}
\affiliation{Indian Institute of Technology, Gandhinagar, Gujarat-382355, India}

\begin{abstract}
\noindent
This work explores the prospect of using the plunge to identify potential black hole mimickers. We show that the plunge excites two generic spectral features. (i) At low frequencies, there is a comb of sharp resonances at the real parts of the mimicker quasi-normal modes. (ii) Above a threshold $M\omega_{\rm th}\!\approx\!0.39$ (for the dominant mode), the spectrum undergoes a qualitative break: with the black hole mimicker displaying significant deviations from the black hole. Though individual plunge SNRs in extreme mass ratio events are low and detecting them in a sea of noise is difficult, the coherent spectral features identified here may allow for enhancing the SNR by using multiple events.
\end{abstract}

\maketitle
\section{Introduction.}
\label{sec:Intro}

Gravitational wave (GW) observations by the LIGO-Virgo-KAGRA (LVK) collaboration \cite{LIGOScientific:2014pky, VIRGO:2014yos, LIGOScientific:2016aoc, LIGOScientific:2018mvr, LIGOScientific:2020ibl, KAGRA:2020tym, KAGRA:2021vkt} could allow us to probe for signatures of new physics with unprecedented accuracy \cite{LIGOScientific:2016lio, LIGOScientific:2018dkp, LIGOScientific:2019fpa, LIGOScientific:2020tif, LIGOScientific:2021sio}. One of the new avenues that the direct detection of GWs has opened is the search for potential black hole mimickers. As an alternative to the black hole hypothesis, various models of horizonless compact objects with a compactness comparable to a black hole are known like gravastars \cite{Mazur:2001fv, Visser:2003ge}, boson stars \cite{Jetzer:1991jr, Schunck:2003kk, Liebling:2012fv}, quantum black holes \cite{Bekenstein:1974jk, Bekenstein:1995ju, Ashtekar:1997yu}, fuzzballs \cite{Mathur:2005zp} and others \cite{Morris:1988tu, Holdom:2016nek, Raposo:2018rjn}. Such compact objects are difficult to distinguish from black holes observationally and could potentially `mimic' a black hole. These black hole mimickers share the feature of perturbations propagating inwards not being completely absorbed, instead being reflected due to the absence of an event horizon. Detection of any such black hole mimicker will be an emissary of new physics, yet to be uncovered.\\

As the compactness of these black hole mimickers are similar to that of black holes, they will reflect GW perturbations on a surface very close to the location of the event horizon of a black hole of comparable mass and spin. In the case of a Schwarzschild-like black hole mimicker, the reflection of perturbations takes place on a surface located very close to $2M$, say, $r_s=2M(1+\epsilon)$ such that $\epsilon\ll1$, even as small as $\mathcal{O}(l_p/M)$. The perturbation at each frequency will be reflected completely or partially, depending on the internal properties of the system. Modifying the boundary condition imposed on perturbations near $2M$ with a black hole exterior is expected to affect the system's response and behaviour, which has pointed towards some key observational tools which could be used to identify such black hole mimickers, this includes their quasinormal mode (QNM) spectrum \cite{Maggio:2020jml, Laghi:2020rgl, Saketh:2024ojw}, tidal deformability \cite{Cardoso:2017cfl, Nair:2022xfm, Chakraborty:2023zed, Berti:2024moe, Silvestrini:2025lbe} and more \cite{Abedi:2016hgu, Cardoso:2017cqb, Mark:2017dnq, LongoMicchi:2019wsh, Datta:2019epe, Datta:2020rvo, Agullo:2020hxe, Chakraborty:2022zlq, Chakravarti:2023wlc, Datta:2024vll, Bhattacharya:2025xko, Bambi:2025wjx}. \\

Among the various domains of exploration into the signatures of potential black hole mimickers, one key avenue is the possibility of using extreme mass ratio inspirals (EMRI). In this context, \cite{ Cardoso:2019nis, Maggio:2021uge} have demonstrated that the gravitational radiation at asymptotic infinity during a quasi-circular inspiral will have characteristic resonances at the real parts of the QNM frequencies of the black hole mimicker. These works have also examined the observational consequences of such resonances on the gravitational waveform during an EMRI and the prospect of identifying them.\\

However, such an analysis using EMRIs can only probe the system up to the GW frequency associated with the innermost stable circular orbit (ISCO). Since, during a quasi-circular inspiral, the system is perturbed by monochromatic GWs with frequencies associated with the circular orbit at each instant. The GW energy radiated away causes it to adiabatically reduce its radius until it reaches the ISCO. Since there are no stable circular orbits smaller than ISCO, the highest frequency that can be probed during the inspiral is approximately the frequency associated with the ISCO, $\omega_{\textrm{ISCO}}$ \cite{Detweiler:1978ge, Poisson:1994yf, Ori:2000zn, OShaughnessy:2002tbu, Sundararajan:2008bw, Barack:2009ey, Hadar:2009ip, Rom:2022uvv}. This inability to probe the system beyond $\omega_{\textrm{ISCO}}$ leaves much on the table \cite{Mukherjee:2025wxa}. One could argue that higher frequency perturbations will be a better probe of a potential reflecting surface near the supposed horizon since such perturbations will not `see' the effective potential and thus be directly reflected by the surface at $r_s$, serving as a direct probe of near-horizon physics.\\

In order to get a handle on how high the perturbing GW frequency should be to probe the near-horizon region directly, we can start by noticing that for a Schwarzschild-like black hole mimicker, when perturbed by the $(\ell,m)=(2,2)$ modes of the metric perturbations, the effective potential will be less than the square of the perturbing frequency when $M\omega\gtrsim 0.39$. This suggests that we can expect some qualitative changes in the behaviour of perturbations when the associated frequency is larger than $\omega_{\text{th}}\sim0.39/M$. Observe, even if the modifications to Schwarzschild-like black hole mimickers arise at $r_s$ such that $\epsilon\sim l_p/M$. We can still expect the qualitative changes in the GW spectrum to arise around the same $\omega_{\text{th}}$, which is several orders of magnitude smaller than the frequency scales associated with $l_p$. This $\omega_{\text{th}}$ will be very close to the real part of the fundamental QNM frequency of a Schwarzschild black hole, a back-of-the-envelope calculation shows that, for a typical supermassive black hole, these frequencies lie well within the frequency range where detectors such as LISA are most sensitive \cite{LISA:2017pwj, LISA:2022kgy, LISA:2024hlh}.\\

Given these observations, we are led to look for new tools that could probe potential black hole mimickers at frequencies beyond $\omega_{\text{th}}$. In this direction, we expect the GW spectrum of the final plunge in a GW event to be a powerful tool to identify potential black hole mimickers. This is because GW radiation during a plunge is not monochromatic, with perturbations across all frequencies being excited \cite{Hadar:2009ip, Rom:2022uvv}, which allows us to probe the system with frequencies beyond $\omega_{\text{th}}$, which could directly explore the region of the spacetime close to the supposed horizon.\\

Based on the above motivations, in this work, we analyse the energy spectrum of the GW radiation emitted during the plunge of a light compact object into a much more massive black hole mimicker. For a GW event, this plunge should start post-ISCO, from the geodesic universal infall (GUI) trajectory \cite{ Ori:2000zn, OShaughnessy:2002tbu, Sundararajan:2008bw, Barack:2009ey, Hadar:2009ip, Rom:2022uvv}. In this work, we will treat a direct plunge from asymptotic infinity as a surrogate model for the GUI plunge and look for potential deviations in the energy spectrum of the emitted GW radiation. Our analysis of the direct plunge demonstrates significant deviation in the energy spectrum at higher frequencies $\omega\gtrsim\omega_{\text{th}}$ in addition to the characteristic resonances at lower frequencies.\\

Such orders of magnitude deviation at higher frequencies, accompanied by the sheer volume of individual observations that could be made possible with next-generation detectors such as LISA \cite{Babak:2017tow}, compounded by the recent developments in using stacking methods \cite{Yang:2017zxs} to enhance the signal-to-noise ratio (SNR) of GW events, suggests that despite the current difficulties with the SNR in the plunge phase of extreme mass ratio (EMR) events, the GW radiation during plunge could plausibly become a powerful tool to identify black hole mimickers over a four years observation run of LISA.\\

We will start in \cref{sec:Back} with an exploration of the theoretical background of perturbations on a Schwarzschild-like black hole mimicker. Here, we will identify general features expected at low and high frequencies in the spectrum of a black hole mimicker. It is followed by \cref{sec:GWSpec}, where we restrict to the direct plunge into a black hole mimicker and describe how to model the plunge. In \cref{sec:Results.}, we will discuss the result of the numerical analysis and the subtleties in interpreting our results. Finally, we will conclude with \cref{sec:Conclusion}.



\section{Theoretical background.}
\label{sec:Back}

\subsection{Perturbations on a Schwarzschild exterior.}
\label{subsec:Plunge_Schwarzschild}

Since this work investigates Schwarzschild-like black hole mimickers, whose exterior can be described using a Schwarzschild metric. The metric perturbations outside its surface $r_s=2M(1+\epsilon)$, is expected to be described by the Regge-Wheeler \cite{Regge:1957td} and  Zerilli \cite{Zerilli:1970se, Zerilli:1970wzz} equations, which can be expressed together as
\begin{equation}
\label{Eq:Pert_Eq}
\left[-\frac{\partial^2}{\partial t^2}+\frac{\partial^2}{\partial x^2}-V_{\ell}^{(\pm)}(r)\right] X_{\ell m}^{(\pm)}(t, r)=S_{\ell m}^{(\pm)}(t, r)~.
\end{equation}
With $X_{\ell m}^{(\pm)}(t, r)$ being the radial part of the perturbations, $V_{\ell}^{(\pm)}(r)$ the associated potential, $S_{\ell m}^{(\pm)}(t, r)$ the source term and $x=r+2M\log(r/2M-1)$ being the tortoise coordinate.  \\

The superscript $(\pm)$ denotes the parity of the perturbation, $(+)$ for polar parity and $(-)$ for axial parity, which corresponds to the Zerilli and Regge-Wheeler modes of the perturbation, respectively. The effective potentials associated with each mode of the perturbation can be expressed as
\begin{equation}
\label{Eq:Potentials}
\begin{aligned}
    V_{\ell}^{(+)}(r)&=\frac{f(r)}{r^2 \Lambda(r)^2}\left[2 \lambda^2(\Lambda(r)+1)+\frac{18 M^2}{r^2}\left(\lambda+\frac{M}{r}\right)\right],\\
    V_{\ell}^{(-)}(r)&=\frac{f(r)}{r^2}\left[\ell(\ell+1)-\frac{6M}{r}\right].
\end{aligned}
\end{equation}
Where we have $f(r)=1-2M/r$, $\lambda=(\ell+2)(\ell-1)/2$ and $\Lambda(r)=\lambda+3M/r$. These equations can be expressed in the Fourier domain as 
\begin{equation}
\label{Eq:Pert_Eq_four}
\left[\frac{\partial^2}{\partial x^2}+\omega^2-V_{\ell}^{(\pm)}(r)\right] X_{\ell m \omega}^{(\pm)}(r)=S_{\ell m \omega}^{(\pm)}(r)~, 
\end{equation}
such that 
\begin{equation}
\label{Eq:timsor}
    \begin{aligned}
    X_{\ell m}^{( \pm)}(t, r)&=\frac{1}{2 \pi} \int_{-\infty}^{\infty} \mathrm{d} \omega e^{-i \omega t} X_{\ell m \omega}^{( \pm)}(r)~,\\
    S_{\ell m}^{( \pm)}(t, r)&=\frac{1}{2 \pi} \int_{-\infty}^{\infty} \mathrm{d} \omega e^{-i \omega t} S_{\ell m \omega}^{( \pm)}(r)~.\\
\end{aligned}
\end{equation}
Formal solutions to \cref{Eq:Pert_Eq_four} can be obtained using its Green's function as follows,
\begin{equation}
\label{Eqn:gensol}
    X^{(\pm)}_{\ell m \omega}(x)=\int_{x_s}^{+\infty}G_{\ell m \omega}^{( \pm)}\left(x, x^{\prime}\right) S^{(\pm)}_{\ell m\omega}(x')dx'~.
\end{equation}
Where $x_s=x(r_s)$, the Green's function $G_{\ell m \omega}^{( \pm)}\left(x, x^{\prime}\right)$, can be expressed as
\begin{equation}
\label{Eq:GreenFun}
\begin{aligned}
G_{\ell m \omega}^{( \pm)}\left(x, x^{\prime}\right)= & \frac{1}{W_{\ell m \omega}^{( \pm)}}\left[X_{\ell m \omega}^{( \pm), \mathrm{up}}(x) X_{\ell m \omega}^{( \pm), \mathrm{in}}\left(x^{\prime}\right) \Theta\left(x-x^{\prime}\right)\right. \\
& \left.+X_{\ell m \omega}^{( \pm), \mathrm{in}}(x) X_{\ell m \omega}^{( \pm), \mathrm{up}}\left(x^{\prime}\right) \Theta\left(x^{\prime}-x\right)\right],
\end{aligned}
\end{equation}
with $X_{\ell m \omega}^{( \pm), \mathrm{in/up}}\left(x^{\prime}\right)$ being linearly independent solutions to the homogeneous part of \cref{Eq:Pert_Eq_four}, that is
\begin{equation}
    \left[\frac{\partial^2}{\partial x^2}+\omega^2-V_{\ell}^{(\pm)}(r)\right] X_{\ell m \omega}^{(\pm),\mathrm{up/in}}(r)=0.
\end{equation}
Note that $W_{\ell m \omega}^{( \pm)}$ is the Wronskian of $X_{\ell m \omega}^{( \pm), \mathrm{in/up}}$.\\ 

If we choose $X_{\ell m \omega}^{( \pm), \mathrm{up/in}}$ with boundary conditions near the surface and asymptotic infinity such that 
\begin{equation}
\label{Eq:HomBound}
\begin{aligned}
X_{\ell m \omega}^{( \pm), \mathrm{up}} &\simeq
\begin{cases}
B_{\ell m \omega}^{( \pm), \mathrm{in}} e^{-i \omega x} + B_{\ell m \omega}^{( \pm), \mathrm{out}} e^{+i \omega x}, & x \rightarrow x_s \\
e^{i \omega x}, & x \rightarrow +\infty
\end{cases}~\\
X_{\ell m \omega}^{( \pm), \mathrm{in}} &\simeq
\begin{cases}
e^{-i \omega x} + R\, e^{i \omega x}, & x \rightarrow x_s \\
A_{\ell m \omega}^{( \pm), \mathrm{in}} e^{-i \omega x} + A_{\ell m \omega}^{( \pm), \mathrm{out}} e^{+i \omega x}, & x \rightarrow +\infty
\end{cases}~.
\end{aligned}
\end{equation}
Using the behaviour of $X_{\ell m \omega}^{( \pm), \mathrm{up/in}}$ at asymptotic infinity, we have the Wronskian to be 
\begin{equation}
\label{Eq:Wronskian}
W_{\ell m \omega}^{( \pm)}=2 i \omega A_{\ell m \omega}^{( \pm), \text {in }}.
\end{equation}

Now, observe that the general solution from \cref{Eqn:gensol} has the following behaviour.
\begin{align}
\label{Eq:Near/far}
\lim_{x \rightarrow x_s} X_{\ell m \omega}^{( \pm)} 
&\sim \frac{e^{-i \omega x} + R\, e^{i \omega x}}{W_{\ell m \omega}^{( \pm)}} 
\int_{x_s}^{\infty} \mathrm{d} x' \, X_{\ell m \omega}^{( \pm), \mathrm{up}} S_{\ell m \omega}^{( \pm)}~, \nonumber \\
\intertext{and}
\lim_{x \rightarrow +\infty} X_{\ell m \omega}^{( \pm)} 
&\sim \frac{e^{+i \omega x}}{W_{\ell m \omega}^{( \pm)}} 
\int_{x_s}^{\infty} \mathrm{d} x' \, X_{\ell m \omega}^{( \pm), \mathrm{in}} S_{\ell m \omega}^{( \pm)}~.
\end{align}
For a Schwarzschild black hole, we have $R=0$ and $x_s\longrightarrow-\infty$ as the perturbations travel into the event horizon at $r=2M$, while for a black hole mimicker, we have $R\neq0$ and $x_s\approx 2M\log(\epsilon)$ as the perturbations get reflected on the surface at $r_s$.\\

We will next explore how the perturbations being reflected on a surface close to $2M$ manifest in the GW energy spectrum.
 
\subsection{Signatures of a black hole mimicker in the energy spectrum.}
\label{subsec:Effect_Boundary}

Here, we will explore how perturbations being reflected on a surface close to $2M$ in a Schwarzschild-like black hole mimicker will manifest in the GW energy spectrum during a plunge.

\subsubsection{Resonances in the energy spectrum at low frequencies.}
\label{subsubsec:Low_freq}

For perturbation on a Schwarzschild background, the energy carried away to infinity through GW radiation per frequency can be expressed as
\begin{equation}
\label{Eq:dEdOmega}
\frac{\mathrm{d} E_{\ell m}}{\mathrm{~d} \omega}=\frac{\omega^2}{64 \pi^2} \frac{(\ell+2)!}{(\ell-2)!}\left[\left|C_{\ell m \omega}^{(+), \text {out }}\right|^2+\frac{4}{\omega^2}\left|C_{\ell m \omega}^{(-), \text {out }}\right|^2\right].
\end{equation}
Where, $C_{\ell m \omega}^{(\pm), \text {out }}$ are the amplitudes of the outgoing modes of the solution $X_{\ell m \omega}^{( \pm)}(r)$ \cite{Regge:1957td, Zerilli:1970se, Zerilli:1970wzz}. One can read off from \cref{Eq:Near/far} that
\begin{equation}
\label{Eq: OutConst}
    C_{\ell m \omega}^{(\pm), \text {out }}=\frac{1}{W_{\ell m \omega}^{( \pm)}} \int_{x_s}^{\infty} \mathrm{d} x^{\prime} X_{\ell m \omega}^{( \pm), \mathrm{in}} S_{\ell m \omega}^{( \pm)}~.
\end{equation}
As black hole mimickers have QNM frequencies with a small imaginary part, and since the QNM frequencies correspond to a vanishing Wronskian, from \cref{Eq:dEdOmega} and \cref{Eq: OutConst} we can see that there will be resonances in the GW energy spectrum whenever the frequency becomes equal to the real part of the QNM frequency of the black hole mimicker. Therefore, the GW energy spectrum will have characteristic resonances at frequencies separated by $\delta\omega= \pi/2M|\log(\epsilon)|$, as was demonstrated in the context of $dE/dt$ during a pseudo-circular inspirals in \cite{Cardoso:2019nis, Maggio:2021uge}.\\


It is interesting to note that the detectability of potential black hole mimickers based on the de-phasing induced by resonances during the inspiral, as in \cite{Cardoso:2019nis, Maggio:2021uge}, depends on parameters like the spin, $\epsilon$ and $R$ of the black hole mimicker. For the de-phasing to be significant, we need a sufficient number of resonances within $\omega_{\textrm{ISCO}}$. Throughout an EMRI, the cumulative effects of all the resonances will add up to a measurable de-phasing, provided there are enough resonant frequencies for the black hole mimicker within $\omega_{\textrm{ISCO}}$.\\

On the other hand, the observational signature of resonances in the GW spectrum during a plunge suffers from its SNR being low in an EMR event \cite{Babak:2017tow}. However, since the resonances are characterised by the QNM frequencies of the primary, we cannot rule out the possibility that one may be able to uncover such coherent features by stacking signals from multiple GW events \cite{Yang:2017zxs}, given the very large number of GW detections made possible with next generation detectors \cite{Babak:2017tow}.\\ 

Besides the resonances at lower frequencies, another more powerful observational tool is unlocked in a plunge, which we shall discuss in the following section.

\subsubsection{New avenues at higher frequencies}
\label{subsubsec:High_freq}

Let us begin by noticing that for the $(\ell,m)=(2,2)$ modes of the metric perturbations in \cref{Eq:Pert_Eq}, something interesting happens when $M\omega\gtrsim0.39$. Above this, we will have the effective potential for the Zerilli modes to be less than $\omega^2$, meaning the waves can transmit more effectively through the potential barrier and probe the near-horizon region. This observation suggests that for larger values of $\omega$, $\omega\gtrsim\omega_{\mathrm{th}}=0.39/M$, we can expect a quantitative change in the energy spectrum for the perturbation. In what follows, we will describe what to expect from a black hole mimicker when $M\omega\gtrsim0.39$.\\

In the case of black holes, this change in the qualitative behaviour manifests as the energy spectrum in this region showing an exponential fall-off \cite{Davis:1971gg, Tashiro:1981ae, radinmod, Sago:2002fe, Berti:2010ce, Silva:2023cer}. It can be seen that
\begin{equation}
    \frac{dE}{d\omega}\approx A_{\text{BH}}e^{-k_{\text{BH}}\omega},\quad\omega\gtrsim\omega_{\text{th}}~
\end{equation}
with a positive $k_{\textrm{BH}}$.\\

Since the perturbations in the exterior of a Schwarzschild-like black hole mimicker are governed by Schrödinger-like wave equations, higher-frequency modes increasingly transmit through the effective potential barrier to reach the surface at $r_s$, get (partially) reflected there, and then transmit back across the barrier to infinity, where they can interfere with the portion scattered by the barrier. Thus, for sufficiently large values of $\omega$ ($\omega\gtrsim\omega_{\mathrm{th}}$), the GW radiation at infinity acquires additional power from the surface-reflected component. So we have
\begin{equation}
\label{Eq:BHM_energy}
\begin{aligned}
\left.\frac{dE}{d\omega}\right|_{\mathrm{mimicker}}
&\approx
\left.\frac{dE}{d\omega}\right|_{\mathrm{BH}} + E^R_\omega,\qquad \omega\gtrsim\omega_{\mathrm{th}}~.
\end{aligned}
\end{equation}
Here, \(\left.{dE}/{d\omega}\right|_{\mathrm{BH}}\) is the energy expected from a black hole \cite{Davis:1971gg, Tashiro:1981ae, Berti:2010ce, Silva:2023cer}, and \(E^R_\omega\) is the contribution from waves that transmitted through the barrier, reflected off the surface at \(r_s\), and transmitted back to infinity, interfering with the scattered field. When \(\omega\) is sufficiently large such that the scattered amplitude is negligible relative to the reflected component, we expect \(E^R_\omega \propto |R|^2\).\\

Observe that such frequencies are usually inaccessible through the dominant modes of the GWs during an inspiral as the highest frequency which can be probed through an inspiral is approximately $\omega_{\mathrm{ISCO}}$ \cite{Detweiler:1978ge, Poisson:1994yf, Ori:2000zn, OShaughnessy:2002tbu, Sundararajan:2008bw, Barack:2009ey, Hadar:2009ip, Rom:2022uvv} and $\omega_{\mathrm{th}}$ is much larger than $\omega_{\mathrm{ISCO}}$. We note here that the above claims regarding the signatures of the absence of a horizon were not dependent on the source term; it only requires the associated GW radiation to be not monochromatic. This is true in the case of a plunge of a point particle into a black hole mimicker from the innermost stable circular orbit (ISCO) \cite{Hadar:2009ip, Rom:2022uvv}, as well as for direct plunge from asymptotic infinity \cite{Zerilli:1970wzz, Davis:1971gg, Tashiro:1981ae, radinmod, Sago:2002fe, Berti:2010ce, Silva:2023cer}.\\

In the following section, we will numerically investigate the $dE/d\omega$ during the direct plunge of a point particle from infinity into a Schwarzschild-like black hole mimicker, treating it as a surrogate model for the plunge from ISCO.

\section{Gravitational spectrum for black hole mimickers.}
\label{sec:GWSpec}
    \begin{figure*}[t!]
    \centering
  \includegraphics[width=\textwidth]{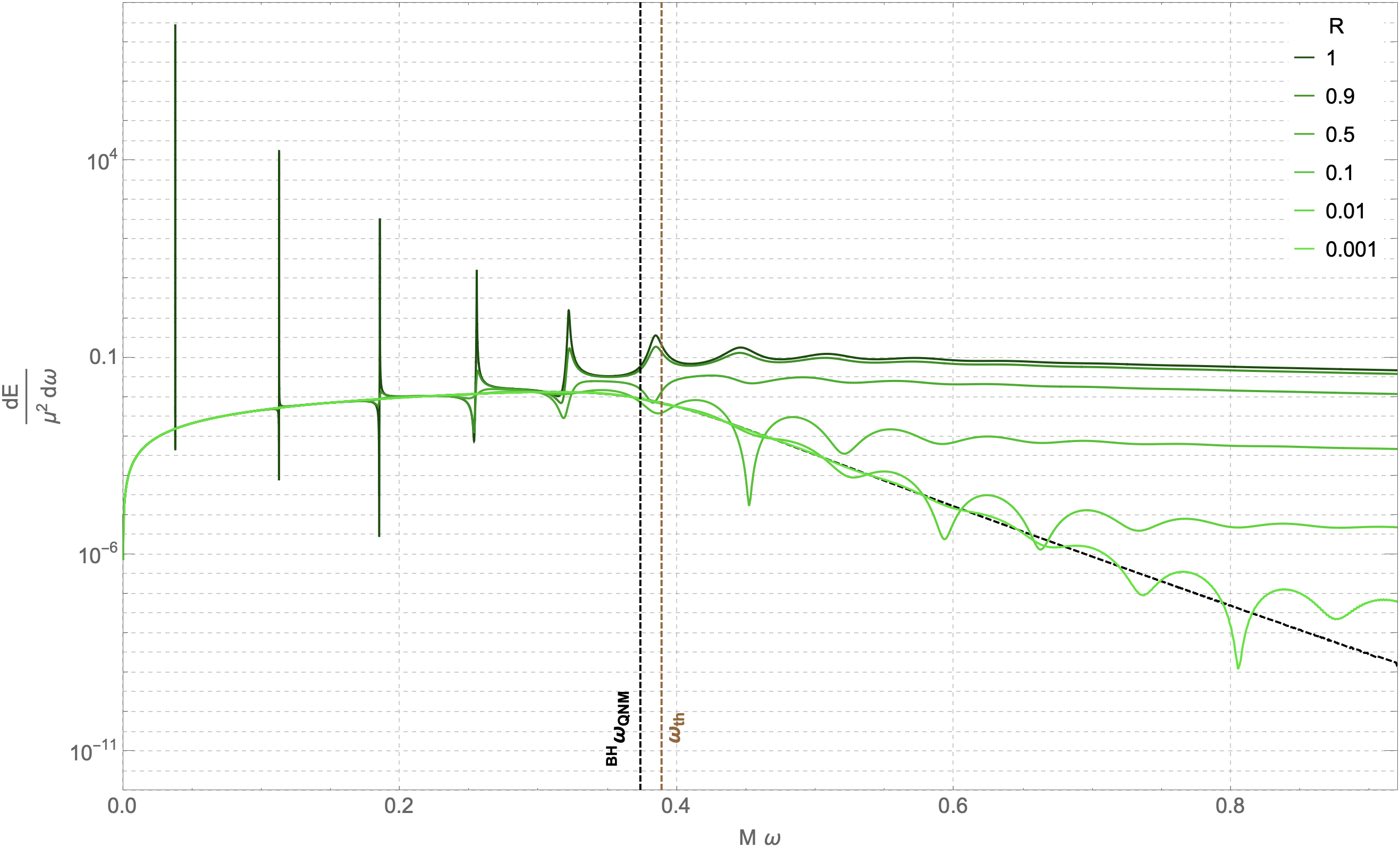}
  \refstepcounter{figure}
\begin{minipage}{\textwidth}
    \justifying
    \noindent
    \textbf{Fig.\ \thefigure:} Plot of $dE/\mu^{2}d\omega$ versus $M\omega$
    for the $(\ell,m)=(2,2)$ modes of the GW perturbations for a black‑hole
    mimicker with a reflecting surface at $r_{s}=2M(1+\epsilon)$,
    $\epsilon=10^{-10}$ from a radial plunge. Here $\mu$ is the mass of the point particle falling
    inwards.  The black and brown vertical lines mark the Schwarzschild fundamental QNM frequency ${}^{\mathrm{BH}}\omega_{\mathrm{QNM}}$ and $\omega_\mathrm{th}$. As expected we see a qualitative change above
    $M\omega_{\mathrm{th}}\gtrsim0.39$ which correctly reproduces the expected behaviour discussed in \cref{subsubsec:High_freq} for different values of $R$. The resonances discussed in \cref{subsubsec:Low_freq} are also correctly reproduced. At low frequencies we recover the characteristic resonances and at high frequencies we can observe a deviation quantified by $A_{\text{m}} e^{-k_{\text{m}}\omega}$, $A_m\propto|R|^2$. Different values of $R$ are represented as varying shades of green, with black hole case being represented by the black dashed line. 
  \end{minipage}
  \label{Fig:R_comparison}
\end{figure*}
\begin{figure*}[t!]
  \centering
  \includegraphics[width=
  \textwidth]{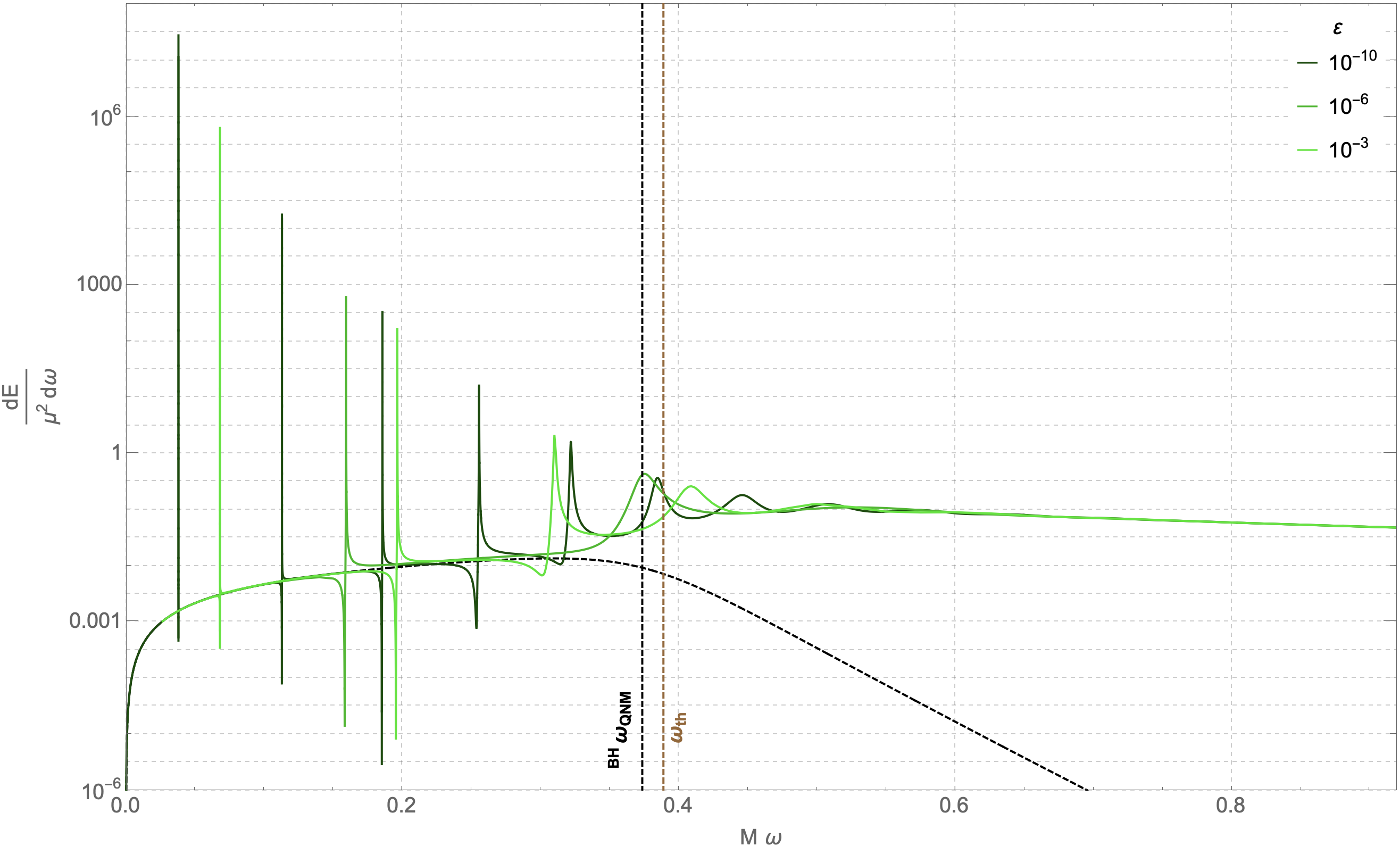}
    \refstepcounter{figure}
\begin{minipage}{\textwidth}
    \justifying
    \noindent
    \textbf{Fig.\ \thefigure:} Plot of $dE/\mu^2d\omega$ against $M\omega$ for the $(\ell,m)=(2,2)$ modes of the GW perturbations for a black hole mimicker with a reflecting surface at $r_s=2M(1+\epsilon)$ from a radial plunge.. Here $\mu$ is the mass of the point particle falling inwards. We have set $R=1$ on the reflecting surface and have varied the separation of the reflecting surface ($\epsilon$) from the supposed horizon of the black hole mimicker. It can be seen that as expected we see a gradual change in the qualitative behaviour of the system above $M\omega_{\text{th}}\gtrsim0.39$. Different values of $\epsilon$ are represented as varying shades of green, with black hole case being represented by the black dashed line. From the plot it can be noticed that the high-frequency behaviour is independent of $\epsilon$.
  \end{minipage}
\label{fig:Newmann_Diff_delta}
\end{figure*}
    \begin{figure*}[t!]
    \centering
  \includegraphics[width=\textwidth]{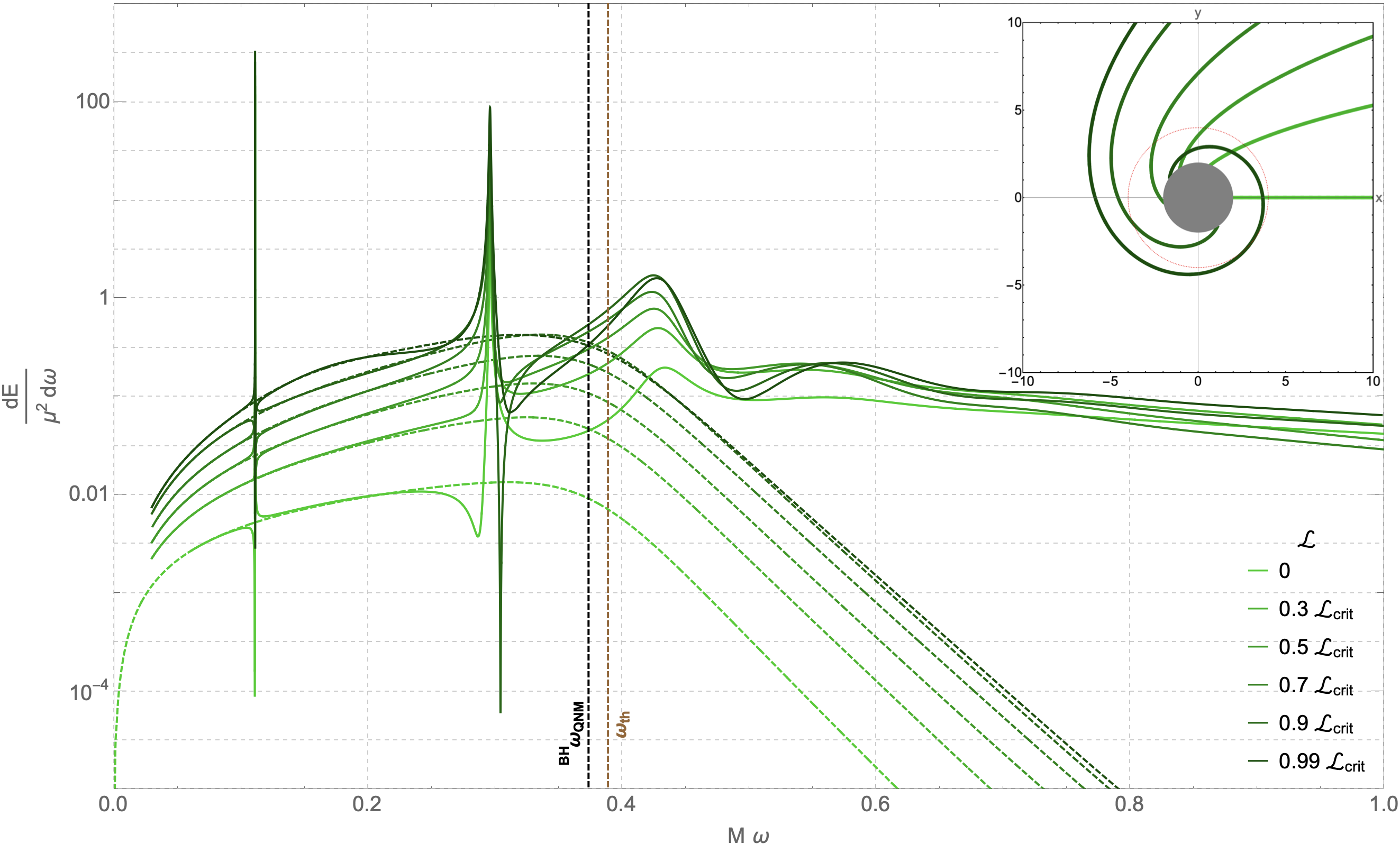}
  \refstepcounter{figure}
\begin{minipage}{\textwidth}
    \justifying
    \noindent
    \textbf{Fig.\ \thefigure:} Plot of $dE/\mu^2d\omega$ against $M\omega$ for the $(\ell,m)=(2,2)$ modes of the GW perturbations for a black hole mimicker with a reflecting surface at $r_s=2M(1+\epsilon)$, $\epsilon=10^{-4}$. We have presented the energy spectrum for point particles with different angular momentum plunging into the black hole mimicker. Different values of $\mathcal{L}$ are represented as varying shades of green, with the corresponding black hole case being represented by the dashed line of the same colour. It can be seen that as expected we see a gradual change in the qualitative behaviour of the system above $M\omega_{\text{th}}\gtrsim0.39$. Here $\mu$ is the mass of the point particle falling inwards.
  \end{minipage}
  \label{Fig:Rotating_energy_flux.pdf}
\end{figure*}
In an EMR event with a primary of mass $M$ and secondary of mass $\mu$, the inspiral ends when the secondary reaches the ISCO, and it is followed by a short transition period where the secondary transitions from an inspiral into a plunge. The final plunge of the secondary into the primary in an EMR event can be well approximated by a geodesic starting just below the ISCO with an angular momentum and energy approximately equal to that of a point mass in orbit at ISCO \cite{ Ori:2000zn, OShaughnessy:2002tbu, Sundararajan:2008bw, Barack:2009ey, Hadar:2009ip, Rom:2022uvv}.\\

A key observation here is that the GW radiation until the ISCO can be computed assuming a pseudo-circular inspiral where the orbital radius slowly decreases as the GW energy radiated away back reacts on the orbit. This results in the GW signal having a slowly changing frequency that grows with time until it reaches the ISCO frequency. However, post-ISCO, the adiabatic approximation used in computing the GW radiation in the inspiral phase is no longer valid. This is because the EMR plunge is characterised by a geodesic motion whose radius is rapidly changing. Thus, the GW radiation during the geodesic plunge is not monochromatic, and it excites GW perturbations across all frequencies, allowing us to go beyond the ISCO frequency.\\

As we noted in the earlier section, higher frequencies open up a new avenue for studying features of the spacetime closer to the surface of the primary. Despite the arguments presented in the earlier section being valid for the plunge from ISCO to the surface of the primary. To get an analytic handle on the plunge that happens post-inspiral in an EMR event, we will work with a surrogate model where we consider the plunge of a secondary with mass $\mu$ from asymptotic infinity into a Schwarzschild-like black hole mimicker of mass $M$, and a reflecting surface at $r_s=2M(1+\epsilon)$ with reflectivity $R$. \\




As in an EMR event $\mu/M\ll1$, modelling the system involves setting the source term $S_{\ell m \omega}^{(\pm)}(r)$ to that of a point particle falling in from asymptotic infinity and evaluating \cref{Eq:dEdOmega}. The source term in turn will depend on the trajectory of the point particle, which can be obtained by solving the geodesic equation, which for motion on the equatorial plane reads
\begin{equation}
\label{Eq:DiffTrag}
\begin{aligned}
\mathrm{d} t_p / \mathrm{d} r_p & =-\left(\mathcal{E} / f(r_p)\right)\left(\mathcal{E}^2-U(r_p)\right)^{-1 / 2}, \\
\mathrm{~d} \phi_p / \mathrm{d} r_p & =-\left(\mathcal{L} / r_p^2\right)\left(\mathcal{E}^2-U(r_p)\right)^{-1 / 2}.
\end{aligned}
\end{equation}
In writing the above equations, we assume that the coordinate location of the point particle is $z^\mu=(t_p,r_p,\pi/2,\phi_p)$ and we have $\mathcal{E}=E/\mu$, $\mathcal{L}=L/\mu$, $U(r)=f(r)(1-\mathcal{L}^2/r^2)$.\\

For the simplest case of a radial infall $(\mathcal{L}=0)$ starting from rest at infinity  $(\mathcal{E}=1)$, $S_{\ell m \omega}^{(-)}$ vanishes and $ S_{\ell m \omega}^{(+)}$ has the following analytic expression \cite{Silva:2023cer, Sago:2002fe}
\begin{equation}
\label{Eq:dirplunso}
\begin{aligned}
    S_{\ell m \omega}^{(+)}(r)&=-8 \pi \mu \mathcal{A}_{\ell m} \frac{f}{r \Lambda}\left[\sqrt{\frac{r}{2 M}}-\frac{2 i}{\omega} \frac{\lambda}{r \Lambda}\right] e^{i \omega t_p(r)}~.
\end{aligned}
\end{equation}
Such that, $\mathcal{A}_{\ell m}=Y_{\ell m}^*(\pi / 2, \phi) \exp (i m \phi)$,  $f(r)=1-2M/r$, $\lambda=(\ell+2)(\ell-1)/2$ and $\Lambda(r)=\lambda+3M/r$ with $Y_{\ell m}$ being the spherical harmonics. $t_p(r)$ can be obtained by solving \cref{Eq:DiffTrag} to read
\begin{equation}
\frac{t_p}{2 M}=-\frac{2}{3}\left(\frac{r}{2 M}\right)^{\frac{3}{2}}-2\left(\frac{r}{2 M}\right)^{\frac{1}{2}}+\log \left[\frac{\sqrt{r /(2 M)}+1}{\sqrt{r /(2 M)}-1}\right]~.
\end{equation}

If we instead consider the case of a non-radial infall ($\mathcal{L}\neq0$) starting from rest ($\mathcal{E}=1$), where the point particle at asymptotic infinity started with a nonzero angular momentum. One has to first numerically compute $t_p(r)$ by solving \cref{Eq:DiffTrag} with appropriate boundary conditions and then plug in the numerical solution into the expression for $S_{\ell m \omega}^{(\pm)}(r)$, with non-zero $\mathcal{L}$ \cite{Sago:2002fe, Silva:2023cer} (see \cref{app:source}), allowing us to get $S_{\ell m \omega}^{(\pm)}(r)$ numerically.\\

In the analysis thus far, we have effectively set the source term in the Fourier domain to be equal to zero when $r_p<r_s$, and when $r_p>r_s$, we are setting it to $S_{\ell m \omega}^{(\pm)}(r)$. This amounts to stopping the time domain analysis at $t_s=t_p(r_s)$; it is known that this could result in corrections in the frequency domain \cite{McKechan:2010kp, Chen:2024ieh} as in real events the source term will not vanish post-$t_s$. This is not a concern in the case of a plunge into a black hole, as here the point particle asymptotes to $2M$ in infinite observer time; while in the context of a black hole mimicker, the time $t_s$ is finite. We will elaborate on this in the upcoming sections.\\ 

With the analytically (when $\mathcal{L}=0$) or numerically (when $\mathcal{L}\neq0$) obtained $S_{\ell m \omega}^{(\pm)}(r)$ in our hands we can go ahead and obtain the GW energy spectrum from such a plunge by computing $dE/d\omega$ using \cref{Eq:dEdOmega} and \cref{Eq: OutConst}. This has to be done numerically by using the methods developed in \cite{radinmod, Silva:2023cer} after imposing appropriate boundary conditions on the GW perturbations at $r_s$. The results of which are discussed in the following section.

\section{Results.}
\label{sec:Results.}

By treating the direct plunge of a point particle as a surrogate for the behaviour of GW radiation from the final GUI plunge \cite{ Ori:2000zn, OShaughnessy:2002tbu, Sundararajan:2008bw, Barack:2009ey, Hadar:2009ip, Rom:2022uvv} in an EMR event, as discussed above. We use the numerical methods developed in  \cite{radinmod, Silva:2023cer} to obtain the GW energy spectrum from a point particle plunging into a Schwarzschild-like black hole mimicker after imposing appropriate boundary conditions on $r_s$. The results of this analysis presented here correctly reproduce the predictions made in \cref{subsec:Effect_Boundary}.\\

In \cref{Fig:R_comparison}, we show the semi-log plot of how $dE/d\omega$ for the $(2,2)$ modes of the GW perturbations emitted during the radial plunge as it changes depending on the reflectivity of the black hole mimicker. Here, we kept $\epsilon=10^{-10}$ and varied $R$ from 1 to $10^{-3}$. In \cref{fig:Newmann_Diff_delta} we compare the spectrum for different values of $\epsilon$, while keeping $R=1$. From both these figures, it can be seen that for lower values of $M\omega$, we correctly regained the resonances as predicted in \cref{subsubsec:Low_freq}, similar to the resonances in $dE/dt$ during inspiral \cite{Cardoso:2019nis, Maggio:2021uge}. Further, we observe that when $\omega\gtrsim\omega_{\mathrm{th}}$, there is a characteristic change in the energy spectrum as predicted in \cref{subsubsec:High_freq}. One can notice that at these higher frequencies the energy spectrum can be described by \cref{Eq:BHM_energy}.\\

We note here that the GW spectrum of these black hole mimickers displays orders of magnitude deviation from the Schwarzschild black hole at $\omega\gtrsim\omega_{th}$. For instance, it can be seen from \cref{Fig:R_comparison} and \cref{fig:Newmann_Diff_delta} that the $dE/d\omega$ for a black hole mimicker with $R=1$, even at the QNM frequency of the Schwarzschild black hole, is ten times that of a Schwarzschild black hole. Further, the difference increases by multiple orders of magnitude at higher frequencies.\\ 

Based on the results of the numerical analysis presented in \cref{Fig:R_comparison} and \cref{fig:Newmann_Diff_delta}, we notice that for sufficiently large $\omega$, for all Schwarzschild-like black hole mimickers the deviations $E^R_\omega$ of \cref{Eq:BHM_energy} can be modelled by
\begin{equation}
\label{Eq:Riff_Correction_Num}
    E^R_\omega= A_{\text{m}} e^{-k_{\text{m}}\omega}, \quad k_m>0,
\end{equation}
with $A_m$ proportional to $|R|^2$. Interestingly, this behaviour appears to be independent\footnote{In our analysis, we have set $R$ to be a constant for simplicity, in general the reflectivity could be a function of $\omega$, in that case the behaviour at higher frequencies will be dictated by $|R(\omega)|^2$.} of location of $r_s$ .\\


Following the analysis with radial plunge ($\mathcal{L}=0$), we proceed to consider plunge scenarios with nonzero values of $\mathcal{L}$. In \cref{Fig:Rotating_energy_flux.pdf}, we display the semi-log plot of the energy spectrum for a point particle plunging with a non-zero angular momentum. We varied the specific angular momentum of the point particle $\mathcal{L}$ from $0$(radial infall) to $99\%$ of the critical angular momentum ($\mathcal{L}_{\text{crit}}=4M$) associated with the marginally bound circular orbit located at $r=4M$. As demonstrated in the subplot within \cref{Fig:Rotating_energy_flux.pdf}, the point particle approached the marginally bound circular orbit when $\mathcal{L}$ approaches $\mathcal{L}_{\text{crit}}$. \\  

We observed that just like the $dE/d\omega$ associated with the radial infall as given in \cref{Fig:R_comparison} and \cref{fig:Newmann_Diff_delta}; for the geodesic plunge with non-zero angular momentum, we regain the characteristic resonances at lower frequencies and orders of magnitude deviations above $\omega_{\mathrm{th}}$. We observe that for sufficiently large values of $\omega$, the spectrum displays a behaviour similar to \cref{Eq:Riff_Correction_Num}.\\

However, from \cref{Fig:Rotating_energy_flux.pdf} we can note that the falloff for different plunge orbits are different as it changes with the specific angular momentum, $\mathcal{L}$ of the point particle, however we observe that the difference is greatly suppressed in relation to the black hole case and the tail is `almost universal'\footnote{We thank Sayak Datta for pointing out that the high-frequency behaviour appears to be less sensitive to $\mathcal{L}$.}. Similar universality has also been reported in \cite{Oshita:2023cjz, Okabayashi:2024qbz}, where the high-frequency GW spectrum becomes largely insensitive to the detailed properties of the source, even in the case of rotating black holes. This universality of the tail against the exact details of the source term further strengthens the case for using a direct plunge as a surrogate for the post-inspiral GUI plunge in studying high-frequency deviations. While the analytic arguments presented in \cref{subsubsec:Low_freq} justifies it at low frequencies.\\


As alluded to in the earlier section, the analysis of the plunge, which we have carried out in the frequency domain, maps to stopping the time domain analysis at $t_p(r_s)$. This could result in corrections in the frequency domain \cite{McKechan:2010kp, Chen:2024ieh}, in relation to real observations where we can expect the energy at each frequency bin to change due to the non-vanishing source term post-$t_s$. However, we note that for the range of frequencies explored in this paper, truncating the source at $t_s$ does not significantly alter the low frequency resonances as well as the spectral excess beyond $\omega_{th}$ (see  \cref{app:extra}). Further, by comparing the results of our analysis with those of greybody factors presented in \cite{Rosato:2025byu}, we conclude that our work captures the correct features of the energy spectrum for the plunge into a black hole mimicker, in the presented frequency range. It should nevertheless be noted that this means a direct time-domain reconstruction, and hence an explicit visualisation of echoes \cite{Abedi:2016hgu, Cardoso:2017cqb} for such black hole mimickers, will require a careful use of windowing methods \cite{McKechan:2010kp, Chen:2024ieh}, as the truncation-induced artefacts in the frequency domain can contaminate the reconstructed signal through the inverse Fourier transform, despite the frequency-domain results presented here remaining robust in the frequency ranges considered.\\

We note that while black hole spectroscopy based on discrete ringdown QNMs is a powerful probe of potential black hole mimickers, the plunge energy spectrum \(dE/d\omega\) carries complementary physical information that may be lost in standard QNM-based ringdown fits. In practice, ringdown analyses model the late-time signal as a superposition of a few damped sinusoids and fit the data in a time window, compressing the response into a small set of complex frequencies and excitation amplitudes, where the fundamental mode dominates. By contrast, the energy spectrum explicitly resolves the full frequency-dependent response of the system to the plunge: it simultaneously captures the low-frequency resonance comb associated with the mimicker QNM spectrum and the qualitative break with an enhanced high-frequency tail for $\omega \gtrsim \omega_{th}$. As illustrated in \cref{subsubsec:High_freq}, this tail represents a continuous scattering process where waves transmit through the photon sphere barrier, reflect off the mimicker surface, and scatter back to infinity; this non-resonant mechanism cannot be efficiently parametrised by a sum of QNMs. This tail instead is related to the graybody factors (see \cref{app:gray}) of the black hole mimicker \cite{Oshita:2023cjz, Okabayashi:2024qbz, Rosato:2025byu}, whose high-frequency behaviour has recently been shown to be stable under small environmental perturbations~\cite{Oshita:2024fzf,Rosato:2024arw}. Thus, the energy spectrum of the GW radiation during plunge can be expected to be a smoking gun observable to identify black hole mimickers.\\

To assess the observational prospects of these plunge spectral features with LISA, let us note that individual EMRI plunges are expected to contribute only a small fraction of the total accumulated SNR, so resolving the resonance comb and the enhanced high-frequency tail in a single event may be challenging. As a rough scaling estimate, for a typical LISA EMRI with an SNR of $\rho_{\rm total}\sim 20$ accumulated over $\sim 10^{4}\text{--}10^{5}$ cycles \cite{Babak:2017tow}. If only a few cycles contribute in the plunge/transition band, a naive duration-based estimate suggests an effective plunge SNR of at most ${\rho_{\rm plunge}\sim \mathcal{O}(10^{-1})}$, with the precise value depending on the transition duration, the source parameters, and the LISA noise curves.\\

However, we note that the spectral features identified for these EMRIs are controlled entirely by the properties of the primary. This opens the possibility of combining information from multiple distinct EMRIs by employing the long inspiral preceding the plunge to determine the plunge time and the central-object parameters well enough to enable coherent alignment of the signals. With coherent alignment, the effective SNR can be expected to improve by $\propto \sqrt{N}$, analogous to \cite{Yang:2017zxs}. If we aim to stack such signals to an effective plunge SNR of $\rho_*\sim5$. It would require us to coherently stack $N \sim (\rho_*/\rho_{\rm plunge})^{2} \sim 10^{3}\text{--}10^{4}$ events. Population studies indicate that LISA may detect anywhere between a few and a few thousand EMRIs per year, subject to astrophysical conditions \cite{Babak:2017tow}, so accumulating sufficient events over a four-year mission duration is plausible.  Nevertheless, for black hole mimickers, accurately modelling the inspiral-to-plunge transition required for enabling phase alignment, necessary for such a coherent stacking, remains an open and challenging problem, as the exact phase evolution during the transition from the inspiral to the plunge will likely deviate from a pure Schwarzschild/Kerr model.\\

We further note that the high-frequency excess above $\omega_{th}$ is a generic feature of black hole mimickers with nonzero reflectivity $R$. For mimickers with detector frame masses in the LISA EMRI range ($M_z \sim 10^{5}\!-\!10^{7} M_\odot$), the threshold frequency $\omega_{th}$ lies within the LISA band, making this feature a particularly promising observational target with such stacking strategies. A quantitative assessment of waveform systematics and stacking strategies for LISA is left to future work.

\section{Conclusion}
\label{sec:Conclusion}

In this work, we investigated the GW spectrum emitted during the plunge of a compact object into a  Schwarzschild-like black hole mimicker with a surface at $r_s=2M(1+\epsilon)$, $\epsilon\ll1$, in an extreme mass ratio event.\\ 

We noted that the plunge could serve as a direct probe for near-horizon features as it excites GW perturbations at frequencies where the effective potential is less than the square of the perturbing frequency. Since the  GW radiation during a direct
plunge is not monochromatic \cite{Sago:2002fe, Silva:2023cer}, just like the post-ISCO GUI plunge in an EMRI \cite{Hadar:2009ip, Rom:2022uvv}. By treating the direct plunge (both radial and non-radial) as a surrogate for the GUI plunge from the ISCO, we studied the GW spectrum at low as well as high frequencies.\\

In the low-frequency limit, we demonstrated the existence of resonances at the real parts of the QNM frequencies of the black hole mimicker. At frequencies beyond $\omega_{\mathrm{th}}$, we demonstrate significant deviations in the energy spectrum. Numerically, we observed that the high-frequency behaviour appears largely insensitive to the location of the surface of the black hole mimicker and the excess energy displays an exponential tail proportional to $|R|^2$.\\


Our work demonstrated that the plunge spectrum contains two
coherent features: (i) a resonance comb below/around $\omega_{\mathrm{th}}=\sqrt{V}|_{\mathrm{max}}$, $V$ being the effective potential, with spacing set by $\epsilon$, and (ii) at high-frequency the spectra break from the black hole case with an additive exponential-tail component. As both the spectral break and the resonance locations are common across events with the same primary, we may be able to use stacking to boost detectability in GUI plunge even if single-event plunge SNRs are low, given the significant number of GW detections made possible with detectors such as LISA.\\

Future research directions include a full time domain analysis of the plunge-merger-ringdown GW waveform for Schwarzschild-like black hole mimickers; the challenge in such an analysis would be modelling the source term after the secondary reached the surface. One could also try repeating our analysis in the frequency domain, but using different windowing methods \cite{McKechan:2010kp, Chen:2024ieh}. There is also the prospect of generalising the study presented here using the GUI plunge source \cite{ Ori:2000zn, OShaughnessy:2002tbu, Sundararajan:2008bw, Barack:2009ey, Hadar:2009ip, Rom:2022uvv} and exploring the possibility of stacking signals from LISA to enhance the SNR post ISCO to identify signatures of such coherent features.

\section*{Acknowledgements}

The authors are thankful to Sudipta Sarkar and Sumanta Chakraborty for extensive discussions. We also thank Elisa Maggio, Rajes Ghosh, Sayak Datta and Shauvik Biswas for sharing detailed comments on the initial draft. We thank IACS, Kolkata, for its hospitality, where part of this work was carried out. This research was supported in part by the International Centre for Theoretical Sciences (ICTS) for participating in the program - Beyond the Horizon: Testing the black hole paradigm (code: ICTS/BTH2025/03). The research of S.N. is supported by the Prime Minister's Research Fellowship (ID-1701653), Government of India. 
\appendix
\crefalias{section}{appendix}
\crefalias{subsection}{appendix}      
\crefalias{subsubsection}{appendix}   


\section{Plunge source terms.}
\label[appendix]{app:source}
Here we will present the explicit expressions for the source term of a point particle plunging on a Schwarzschild background on the equatorial plane with angular momentum $L$, energy $E$, and mass $\mu$ \cite{Sago:2002fe, Silva:2023cer}. For the polar or Zerilli modes of the perturbation, it reads

\begin{equation}
\begin{aligned}
S_{\ell m \omega}^{(+)}= & -i f \frac{\mathrm{~d}}{\mathrm{~d} r}\left[\frac{f^2}{\Lambda}\left(\frac{i r}{f} \tilde{C}_{1 \ell m \omega}+\tilde{C}_{2 \ell m \omega}\right)\right] \\
& +i \frac{f^2}{r \Lambda^2}\left[i \frac{\lambda r^2-3 \lambda M r-3 M^2}{r f} \tilde{C}_{1 \ell m \omega}\right. \\
& \left.+\frac{\lambda(\lambda+1) r^2+3 \lambda M r+6 M^2}{r^2} \tilde{C}_{2 \ell m \omega}\right],
\end{aligned}
\end{equation}
such that $f=1-2M/r$, $\lambda=(\ell+2)(\ell-1)/2$,
\begin{equation}
\begin{aligned}
\Lambda= &\lambda+\frac{3M}{r},\\
\tilde{B}_{\ell m \omega}= & \frac{8 \pi r^2 f}{\Lambda}\left[A_{\ell m \omega}+\frac{1}{\sqrt{\ell(\ell+1) / 2}} B_{\ell m \omega}\right] \\
& -4 \pi \frac{\sqrt{2}}{\Lambda} \frac{M}{\omega} A_{\ell m \omega}^{(1)}, \\
\tilde{C}_{1 \ell m \omega}= & \frac{8 \pi}{\sqrt{2} \omega} A_{\ell m \omega}^{(1)}+\frac{1}{r} \tilde{B}_{\ell m \omega} \\
& -16 \pi r\left[\frac{1}{2} \frac{(\ell+2)!}{(\ell-2)!}\right]^{-\frac{1}{2}} F_{\ell m \omega}, \\
\tilde{C}_{2 \ell m \omega}= & \frac{8 \pi i}{\omega \sqrt{\ell(\ell+1) / 2}} \frac{r}{f} B_{\ell m \omega}^{(0)}-\frac{i}{f} \tilde{B}_{\ell m \omega} \\
& +\frac{16 \pi i r^2}{f}\left[\frac{1}{2} \frac{(\ell+2)!}{(\ell-2)!}\right]^{-\frac{1}{2}} F_{\ell m \omega}, 
\end{aligned}
\end{equation}

\begin{equation}
\begin{aligned}
A_{\ell m \omega} & =\mu \frac{V}{r^2 f^2} Y_{\ell m}^* e^{i \omega t_p} \\
A_{\ell m \omega}^{(1)} & =-i \sqrt{2} \mu \frac{\mathcal{E}}{r^2 f} Y_{\ell m}^* e^{i \omega t_p} \\
B_{\ell m \omega}^{(0)} & =i \mu \frac{\mathcal{E} \mathcal{L}}{V r^3} \frac{1}{\sqrt{\ell(\ell+1) / 2}} \partial_\phi Y_{\ell m}^* e^{i \omega t_p} \\
B_{\ell m \omega} & =-\mu \frac{\mathcal{L}}{r^3 f} \frac{1}{\sqrt{\ell(\ell+1) / 2}} \partial_\phi Y_{\ell m}^* e^{i \omega t_p} \\
F_{\ell m \omega} & =\mu \frac{\mathcal{L}^2}{V r^4}\left[\frac{1}{2} \frac{(\ell+2)!}{(\ell-2)!}\right]^{-\frac{1}{2}} \partial_{\phi \phi} Y_{\ell m}^* e^{i \omega t_p},
\end{aligned}
\end{equation}
$\mathcal{E}=E/\mu$, $\mathcal{L}=L/\mu$, $V=\sqrt{\mathcal{E}^2-U}$, $U=f(1-\mathcal{L}^2/r^2)$ and $t_p$ is the solution to \cref{Eq:DiffTrag}. Here $Y_{\ell m}$ is the spherical harmonics with $\theta=\pi/2$.\\

For the axial or Regge-Wheeler modes of the perturbation, we have,
\begin{equation}
\begin{aligned}
S_{\ell m \omega}^{(-)}= & \frac{8 \pi i f}{r}\left[\frac{1}{2} \frac{(\ell+2)!}{(\ell-2)!}\right]^{-\frac{1}{2}}\left[-r^2 \frac{\mathrm{~d}}{\mathrm{~d} r}\left(f D_{\ell m \omega}\right)\right. \\
& \left.+\sqrt{2 \lambda} r f Q_{\ell m \omega}\right],
\end{aligned}
\end{equation}

\begin{equation}
\begin{aligned}
D_{\ell m \omega} & =i \mu \frac{\mathcal{L}^2}{V r^4}\left[\frac{1}{2} \frac{(\ell+2)!}{(\ell-2)!}\right]^{-\frac{1}{2}} X_{\ell m}^* e^{i \omega t_p}, \\
Q_{\ell m \omega} & =-i \mu \frac{\mathcal{L}}{f r^3} \frac{1}{\sqrt{\ell(\ell+1) / 2}} \partial_\theta Y_{\ell m}^* e^{i \omega t_p},
\end{aligned}
\end{equation}
\begin{equation}
X_{\ell m}=2 \partial_\phi\left(\partial_\theta-\cot \theta\right) Y_{\ell m} .
\end{equation}

\section{Effects of full source term.}
\label[appendix]{app:extra}
Let us assume that the observed black hole mimicker has a source term that includes information about what happens after the secondary reaches $r_s$, we can express it in the time domain as $\mathcal{S}_{\ell m}(t)$. In our model, we restricted ourselves to the plunge, and thus we have set the source term post $t_s=t_p(r_s)$ to zero. So what we have done is our analysis of the plunge is set $S_{\ell m}^{(\pm)}(t)\approx\Theta(t_s-t)\mathcal{S}_{\ell m}(t)$, with $S^{(\pm)}_{\ell m}(t)$ being the plunge source term (as in \cref{Eq:timsor} and \cref{Eq:dirplunso}) up to $t_s$ .\\

We wish to quantify how much the full source term for the black hole mimicker in the Fourier domain $\mathcal{S}_{\ell m\omega}=\int_{-\infty}^{\infty}\mathcal{S}_{\ell m}(t)e^{i\omega t}\,dt$ deviates from the model source therm, which we have used $S_{\ell m\omega}=\int_{-\infty}^{t_s}{S}_{\ell m}(t)e^{i\omega t}\,dt$. So we define the deviation $\Delta_{\ell m}(\omega)$ as
\begin{equation}
\label{Eq:gendif}
    \begin{aligned}
        \Delta_{\ell m}(\omega)&= S_{\ell m \omega}-\mathcal{S}_{\ell m\omega}\\
        &=-\!\int_{t_s}^{\infty}\mathcal{S}_{\ell m}(t)e^{i\omega t}\,dt~.
    \end{aligned}
\end{equation}
Since the time scale of the black hole mimicker will be $\mathcal{O}(M)$, when $M\omega\ll1$, we have
\begin{equation}
\label{Eqn: NumCorr}
    \Delta(\omega)=-\sum_{n=0}^{\infty}\frac{(i\omega)^n}{n!}\,\mu_n,~\mu_n=\int_{t_s}^{\infty} t^n \mathcal{S}_{\ell m}(t)\,dt.
\end{equation}
The above corrections will manifest in the observed spectrum as corrections of the form $\Sigma_i c_i\omega^i$, with each $c_i$ being determined by the $\mu_n$ and thus $\mathcal{S}_{\ell m}(t)$ post merger.\\

One could also show that such a hard cut-off also yields a correction at $M\omega\gg1$ (Which is much larger than the largest value of $M\omega$ considered in this work.) of the form
\begin{equation}
    \Delta_{\ell m}(\omega)=e^{i\omega t_s}\!\left[\frac{\mathcal{S}_{\ell m}(t_s)}{i\omega}-\frac{\mathcal{S}'_{\ell m}(t_s)}{(i\omega)^2}+\cdots\right].
\end{equation}
The above expression can be obtained by repeatedly performing integration by parts on \cref{Eq:gendif}. In practice, this will alter the falloff of the asymptotic tail of $C_{\ell m \omega}^{(\pm), \text {out }}$, making the numerical integration to recover the time domain waveform of such black hole mimickers more difficult.\\

To assess whether the truncation at $t_s$ could meaningfully affect the spectral excess beyond $\omega_{th}$ discussed in the main text, we next provide an order of magnitude estimate of the induced correction at a representative point, $M\omega\simeq 0.7>\omega_{th}$.\\

\section*{Order of magnitude estimate of truncation-induced corrections at a representative frequency}
An estimate of the order of magnitude contribution to the observed GW energy spectrum from the omitted Fourier contribution $\Delta_{\ell m}(\omega)$ follows from the expectation that the post-$t_s$ source persists over a characteristic duration $T_{\rm post}$ of order the primary scale, i.e. $T_{\rm post}=\mathcal{O}(M)$. Further since the time domain source $\mathcal{S}_{\ell m}(t)$ is sourced by a body of mass $\mu$, we can expect $\mathcal{S}_{\ell m}(t) \sim \mu$, we can therefore parameterise
\begin{equation}
|\Delta_{\ell m}(\omega)| \sim \eta\,\mu\,T_{\rm post},
\qquad \eta=\mathcal{O}(1),
\label{Eq:Delta_eta}
\end{equation}
where $\eta$ absorbs the dependence on the detailed post-$t_s$ source profile.\\

To estimate the change in the energy spectrum by incorporating $\Delta_{\ell m}(\omega)$, we can start by noticing that the RW/Zerilli equations are linear. As a result, the outgoing frequency-domain amplitudes are linear functionals of the source (cf. \cref{Eq: OutConst}). So the truncation induces a shift
\begin{equation}
C^{(\pm){\rm, out}}_{\ell m\omega}\ \longrightarrow\ C^{(\pm){\rm, out}}_{\ell m\omega}+\delta C^{(\pm){\rm, out}}_{\ell m\omega},\,
\delta C^{(\pm){\rm, out}}_{\ell m\omega}\sim \kappa\,\Delta_{\ell m}(\omega),
\label{Eq:deltaC_kappa}
\end{equation}
with $\kappa=\mathcal{O}(1)$, encoding the effects from the integral over the homogeneous solution.\\

For a given mode, RW/Zerilli( or the corresponding effective amplitude combination entering \cref{Eq:dEdOmega}), the spectrum is quadratic in the complex amplitude,
\begin{equation}
\frac{dE}{d\omega}=\mathcal{K}(\omega)\,|C^{(\pm){\rm out}}_{\ell m\omega}|^2,
\label{Eq:spectrum_quadratic_B}
\end{equation}
where $\mathcal{K}(\omega)$ is the known real prefactor as infered from \cref{Eq:dEdOmega}. Hence the spectrum after incorporating $\Delta_{\ell m}(\omega)$ is
\begin{equation}
\begin{aligned}
\frac{1}{\mu^2}\frac{d\mathcal{\tilde{E}}}{d\omega}
&=\frac{\mathcal{K}(\omega)}{\mu^2}\,\bigl|C^{(\pm){\rm out}}_{\ell m\omega}+\delta C^{(\pm){\rm out)}}_{\ell m\omega}\bigr|^2 \\
&=\frac{1}{\mu^2}\frac{dE}{d\omega}\left(1+2\alpha\cos\phi+\alpha^2\right),
\label{Eq:Correc_In_r}
\end{aligned}
\end{equation}
where we introduced
\begin{equation}
\alpha\equiv \frac{|\delta C^{(\pm){\rm out}}_{\ell m\omega}|}{|C^{(\pm){\rm out}}_{\ell m\omega}|},
\qquad
\phi\equiv \arg(\delta C^{(\pm){\rm out}}_{\ell m\omega})-\arg(C^{(\pm){\rm out}}_{\ell m\omega}).
\label{Eq:rphi_def}
\end{equation}
Equation~\eqref{Eq:Correc_In_r} makes the leading truncation effect explicit and shows it to be an interference correction ($\propto \alpha\cos\phi$), whose sign is controlled by the unknown relative phase $\phi$ of the omitted contribution.\\

Based on the above discussion, we can now estimate the corrected energy flux at a representative point, say $M\omega\simeq 0.7$, from \cref{Fig:Rotating_energy_flux.pdf}.  For the dominant $(\ell,m)=(2,2)$ contribution, we define the effective amplitude
\(\mathcal{C}_{22\omega}\) through \cref{Eq:dEdOmega} so that
\begin{equation}
\frac{dE_{22}}{d\omega}=\frac{3\omega^2}{8\pi^2}\,\mathcal{C}_{22\omega}^2
\quad\Rightarrow\quad
\frac{\mathcal{C}_{22\omega}}{\mu}
=
\sqrt{\frac{8\pi^2}{3\omega^2}\left(\frac{1}{\mu^2}\frac{dE_{22}}{d\omega}\right)},
\label{Eq:Ceff_from_spectrum_B}
\end{equation}
    \begin{figure*}[t!]
    \centering
  \includegraphics[width=\textwidth]{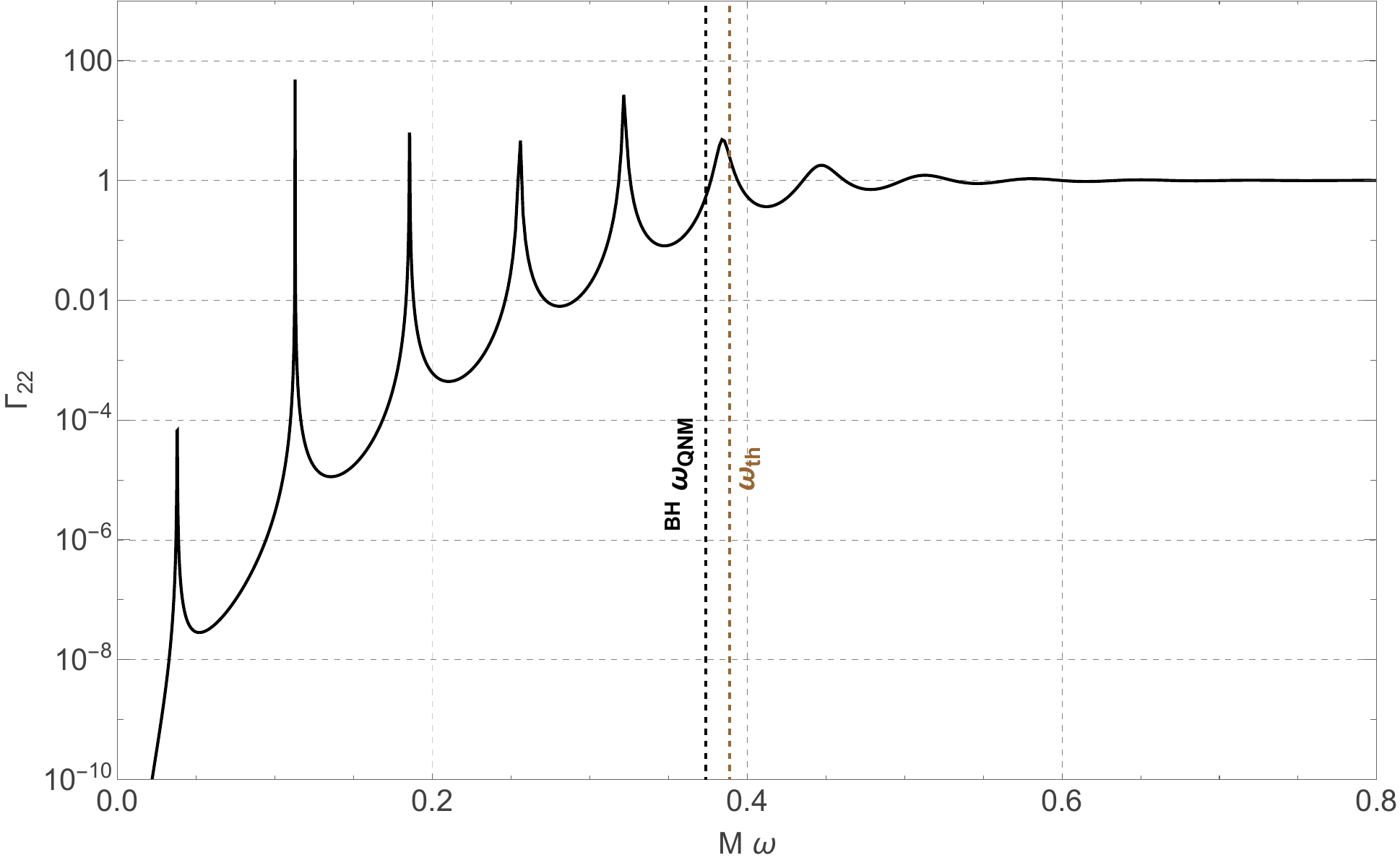}
  \refstepcounter{figure}
\begin{minipage}{\textwidth}
    \justifying
    \noindent
    \textbf{Fig.\ \thefigure:} Plot of the graybody factor against $M\omega$ for the $(\ell,m)=(2,2)$ modes of the GW perturbations for a black-hole mimicker with a reflecting surface at $r_s = 2M(1+\epsilon)$, with $R=1$ and $\epsilon=10^{-10}$. From the figure, it is clear that the graybody factor defined by \cref{Eq:gbfactor} approaches $1$ above $M\omega_{th}\gtrsim0.39$, indicating transmission through the photon sphere. At low frequencies, we see sharp resonances associated with the real parts of the QNMs of the black hole mimicker, consistent with the plunge spectrum analysis in \cref{subsubsec:Low_freq} of the main text and the discussion in \cref{app:gray}.
  \end{minipage}
  \label{Fig:gb}
\end{figure*}
where $dE_{22}/d\omega$ is our numerically computed energy flux from the main text, where the contribution to the source post-$t_s$ was ignored. From the numerical result presented in \cref{Fig:Rotating_energy_flux.pdf}, it can be seen that $(1/\mu^2)\,dE_{22}/d\omega\sim 10^{-1}$ when $M\omega\simeq 0.7$ for all presented values of $\mathcal{L}$. So we find
\begin{equation}
\frac{\mathcal{C}_{22\omega}}{\mu}\simeq 2.32\,M.
\label{Eq:Ceff_numeric_B}
\end{equation}
We wish to now use \cref{Eq:Correc_In_r} to estimate the energy flux if we were to include the post-$t_s$ source term at $M\omega\simeq 0.7$. For this, we will first plug in \cref{Eq:Delta_eta,Eq:deltaC_kappa,Eq:Ceff_numeric_B} into the definition of $\alpha$ from \cref{Eq:rphi_def}. So, for $(\ell,m)=(2,2)$, we have
\begin{equation}
\alpha \sim \frac{|\delta \mathcal{C}_{22\omega}|}{\mathcal{C}_{22\omega}}
\sim \frac{\tilde{\eta}\,T_{\rm post}}{2.32\,M},
\qquad \tilde{\eta}\equiv \eta\kappa=\mathcal{O}(1),
\label{Eq:r_estimate_B}
\end{equation}
In particular, taking $\tilde{\eta}\sim 1$ and $T_{\rm post}\sim M$ yields $\alpha\simeq 0.43$.\\

Thus, we can get an estimate of the energy flux at  $M\omega\simeq 0.7$ after accounting for the truncation effect using \cref{Eq:Correc_In_r}, giving
\begin{equation}
\frac{1}{\mu^2}\frac{d\mathcal{\tilde{E}}}{d\omega}
\approx 0.1\,(1.19+0.86\cos\phi)
\in \left[3.3\times 10^{-2},\,2.0\times 10^{-1}\right].
\label{Eq:B3_numeric_range}
\end{equation}

Thus, under the minimal assumption that the post-$t_s$ source persists for a duration $T_{\rm post}=\mathcal{O}(M)$, the truncation-induced uncertainty in the spectrum at $M\omega\simeq 0.7$ (cf.\ \cref{Fig:Rotating_energy_flux.pdf}) is bounded to be at most a factor of a few; see \cref{Eq:B3_numeric_range}. Consequently, by referring to \cref{Fig:Rotating_energy_flux.pdf}, even after accounting for these truncation effects, the energy spectrum of the black-hole mimicker at $M\omega\simeq 0.7$ remains orders of magnitude larger than the corresponding Schwarzschild black-hole spectrum at the same $M\omega$.\\

In summary, the exact truncation-induced modification to the energy spectrum cannot be determined without specifying the complete time-domain source $S_{\ell m}(t)$ for $t>t_s$, since both the magnitude and the relative phase of the omitted Fourier contribution $\Delta_{\ell m}(\omega)$ depend on the (unknown) post-$t_s$ dynamics. Nevertheless, adopting the minimal assumption that the post-$t_s$ source persists for a characteristic duration $T_{\rm post}\sim M$, we were able to estimate the corresponding correction and explicitly bound the range of values that the corrected spectrum $(1/\mu^2)\,d\mathcal{\tilde{E}}/d\omega$ can take at a representative point $M\omega\simeq 0.7>\omega_{th}$; see \cref{Eq:B3_numeric_range}. A quick glance at \cref{Fig:Rotating_energy_flux.pdf} and \cref{Eq:B3_numeric_range} demonstrates that in the frequency ranges studied in this paper, even after accounting for truncation effects, the spectral excess above $\omega_{th}$ discussed in the main text remains robust and stays several orders of magnitude larger than the corresponding Schwarzschild black-hole spectrum. Thus, we can conclude that the high frequency enhancements presented in the paper demonstrate a qualitative change in the behaviour of the system beyond $\omega_{th}$, robust against the details of the post-$t_s$ source and the associated windowing effects.\\

\section{Graybody factors of a black hole mimicker and the energy spectrum.}
\label[appendix]{app:gray}
In black hole perturbation theory, the greybody factor provides a quantitative measure of how efficiently perturbations transmit through the photon sphere, i.e. the peak of the effective Regge-Wheeler/Zerilli potential.\\

Graybody factors are defined from the homogeneous perturbations on the background spacetime. For each mode $(\ell,m)$ and frequency $\omega$, the radial master equation is a one-dimensional Schrödinger-like equation (cf. \cref{Eq:Pert_Eq}),
and homogeneous solutions at asymptotic infinity admit a standard scattering decomposition. For instance let us concentrate on the homogenous solution $X_{\ell m \omega}^{( \pm), \mathrm{in}}$ defined by \cref{Eq:HomBound}, with
\begin{equation}
X_{\ell m \omega}^{( \pm), \mathrm{in}}\xrightarrow[r_*\to+\infty]{}\;A^{(\pm){\rm, in}}_{\ell m\omega}e^{-i\omega r_*}+A^{(\pm){\rm, out}}_{\ell m\omega}e^{+i\omega r_*},
\end{equation}
where $A^{\rm in/out}_{\ell m\omega}$ are the ingoing/outgoing amplitudes at asymptotic infinity. The potential barrier induces partial reflection of such perturbations, characterised by a notion of transmittivity also referred to as the greybody factor,
\begin{equation}
\label{Eq:gbfactor}
\Gamma_{\ell m}(\omega)=\frac{1}{|A^{(\pm){\rm in}}_{\ell m\omega}|^2}.
\end{equation}
For a black hole mimicker, the above-defined quantity should be interpreted as a photon-sphere transmissivity, not a true absorption probability (more details regarding the subtleties in the notion of graybody factors for black hole mimickers can be found in \cite{Rosato:2025byu}).\\

Now, if for $\omega>\omega_{th}$ the associated perturbations are able to probe the near-horizon region more effectively, we expect the graybody factor defined through \cref{Eq:gbfactor} to approach $1$ when $\omega\gtrsim\omega_{th}$, in fact, this is precisely what we observe in \cref{Fig:gb}\\

We further note that this notion of graybody factor is not merely a convenient scattering diagnostic; it also enters directly as an overall prefactor in the radiated energy flux measured at asymptotic infinity and thus will carry the same low-frequency resonance comb in $dE/d\omega$ as discussed in the main body. To see this one can use \cref{Eq:dEdOmega,Eq: OutConst} along with the expression for the Wronskian between $X_{\ell m \omega}^{( \pm), \mathrm{in}}$ and $X_{\ell m \omega}^{( \pm), \mathrm{up}}$ from \cref{Eq:Wronskian} to obtain $dE/d\omega \sim \Gamma_{\ell m}(\omega)\times {\rm (inegral\, over\, source)}$. Thus, we can analytically expect the low-frequency resonance comb associated with the mimicker QNM spectrum demonstrated in \cref{subsubsec:Low_freq} to appear in the graybody factor plot as well, consistent with \cref{Fig:gb}.


\end{document}